

\input harvmac

\def\su33{$SU(3)^3$}

\Title{HUTP-92/A068} {The Breaking of the \su33 Gauge Group}
\centerline{Meng Y. Wang}
\centerline{Eric D. Carlson}
\bigskip\centerline{Lyman Laboratory of Physics}
\centerline{Harvard University}\centerline{Cambridge, MA 02138}
\medskip

\vskip.3in
We discuss why the \su33 supersymmetric model with the most general
superpotential can naturally break to the standard model if gauge
singlets and a discrete symmetry are included. This mechanism does away with
the need for fine-tuning in the form of the assumed absence of certain terms
in the superpotential. It also automatically guarantees that any abelian
discrete phase symmetry of the GUT will survive the symmetry breaking. Such a
discrete symmetry, also known as the matter parity, is needed to suppress both
proton decay\  \ref\rmatterp{M.C. Bento, L. Hall and G.G. Ross, {\it Nucl.
Phys. }{\bf B292} (1987) 400.}\ and the flavor changing neutral current
(FCNC), and may help solve the hierarchy problem.

\Date{12/92}

\newsec{Introduction}

Previous work\ \ref\rlthiggs{M.Y. Wang and E.D. Carlson, Harvard Preprint
HUTP-92/A062.}\ has indicated that \su33 has many attractive features as a
high-energy supersymmetric gauge group. Besides being a maximal subgroup of
$E_6$, which may arise naturally in string theories\ \ref\roxford{B. Greene,
K.H. Kirklin, P.J. Miron and G.G. Ross, {\it Phys. Lett. }{\bf B180} (1986) 69
; {\it Nucl. Phys. }{\bf B278} (1986) 667 ; {\bf B292} (1987) 606.}, it is
possible to use discrete symmetries to naturally allow the conventional Higgs
doublets to be light. In this paper we will explore this idea further,
focusing particularly on the gauge group breaking.

The gauge group in our earlier paper is based on the group $SU(3)_C \times
SU(3)_L \times SU(3)_R$, where $SU(3)_C$ is the familiar color $SU(3)$,
$SU(3)_L$ contains weak $SU(2)$, and $SU(3)_R$ contains the right-handed
analog of weak $SU(2)$. This group is one of the maximal subgroups of $E_6$,
with the fundamental 27--dimensional representation of $E_6$ becoming a direct
sum of three irreducible representations under \su33\ : $\Psi_L\ :\
(3,\bar{3},1),\ \Psi_R\ :\ (\bar{3},1,3),\ \Psi_\ell\ :\ (1,3,\bar{3})$,
corresponding to the quarks, the anti-quarks, and the leptons respectively.
The explicit assignment of left-handed particles is as follows:
$$\eqalign{\matrix{\Psi_L\cr (3,\bar{3},1)\cr}\
    &:\ \left(\matrix{u&d&B\cr u&d&B\cr u&d&B\cr}\right),\cr
   \matrix{\Psi_R\cr (\bar{3},1,3)\cr}\
    &:\ \left(\matrix{u^*&u^*&u^*\cr d^*&d^*&d^*\cr B^*&B^*&B^*\cr}\right),\cr
   \matrix{\Psi_\ell\cr (1,3,\bar{3})\cr}\
    &:\ \left(\matrix{E^o&E^-&e^-\cr E^+&{E^o}^*&\nu\cr
      e^+&\nu_R&N^o\cr}\right),\cr}$$
where $B$ is an additional superheavy down-type quark, $B^*$ is its
anti-particle, and $E$'s and $N^o$ are new superheavy leptons.

The Higgs needed to break\ \su33\ to the standard model
can be put into a $(1,3,\bar{3})$ representation together with those needed to
break weak $SU(2)$. In supersymmetrized theories, they are just additional
generations of leptons. The VEV's which break $SU(3)^3$ are usually written as
\eqn\evev{\left(\matrix{0&0&0\cr 0&0&0\cr 0&0&v\cr}\right)\ \hbox{\rm and}\
\left(\matrix{0&0&0\cr 0&0&0\cr 0&w&0\cr}\right).}
We then impose an additional discrete symmetry which helps explain the
hierarchy problem\rlthiggs. However, two assumptions in Ref.\
\rlthiggs\ have not been justified: namely, why only two of the superfields
develop vacuum expectation values (VEV's), and why the VEV's of the mirror
superfields do not break the discrete symmetry. Conventional wisdom does not
provide a satisfactory answer to these questions. In fact, they contain other
intrinsically unpleasant features. We therefore are forced to take a new, more
careful look at the gauge symmetry breaking mechanism.

In Section 2, we first introduce the conventional symmetry breaking mechanism,
emphasizing how it fails to explain a few key questions. In Section 3, we
detail our mechanism, showing that it is phenomenologically feasible. In
Section 4, we summarize our work.

\newsec{The Old Breaking Mechanism of \su33}

The conventional method of generating the two necessary VEV's, as explained in
detail by the authors of Ref.\ \ref\rnath{P. Nath and R. Arnowitt, {\it Phys.
Rev.} {\bf D39} (1989) 2006} is based on one specific string inspired model\
\roxford. As is characteristic of string inspired models, the
renormalizable part of the superpotential contains only trilinear terms,
$\Psi\Psi\Psi$. Thus the VEV's in\ \evev\ are the most general $F$-flat
direction of the superpotential after a choice of basis if we assume only two
multiplets grow VEV's. Unfortunately, the $D$-flatness condition is not
satisfied by them. To make the symmetry breaking possible, it is necessary
to introduce the ``mirror particles", which come from the $\overline{27}$
representation of $E_6$. They transform under\ \su33\ like
$\bar\Psi_L\ :\
(\bar{3},3,1),\ \bar \Psi_R\ :\ (3,1,\bar{3}),\ \bar\Psi_\ell\ :\
(1,\bar{3},3).$ If the $(1,\bar{3},3)$ parts gain the following VEV's,

$$\left(\matrix{0&0&0\cr 0&0&0\cr 0&0&v^*\cr}\right)\ \hbox{\rm and}\
\left(\matrix{0&0&0\cr 0&0&w^*\cr 0&0&0\cr}\right),$$
both $F$- and $D$-flatness will be satisfied. The number of light
generations $N_g$ will be equal to the difference between the number of
supermultiplets and that of their mirror partners, {\it i.e.}
\eqn\enumbergen{3\ =\ N_g\ =\ (\#\ of\ \Psi_x)\ -\ (\#\ of\ \bar\Psi_x),}
where $x=C,L,R$. Eq.\ \enumbergen\ automatically guarantees that the model is
free of anomalies.

However, these conventional models have several difficulties. One has to do
with the magnitudes of $v$ and $w$, which are undetermined until we include
non-renormalizable and soft SUSY breaking terms. In order to have
phenomenologically acceptable values for the VEV's, {\it i.e.}
$v,w \ge 10^{16}GeV$\ \rlthiggs\ \ref\rrunning{W. de Boer and H. F{\"
u}rstenau, {\it Phys. Lett. }{\bf B260} (1991) 447.}, the first two leading
non-renormalizable terms have to vanish\ \rnath. So far, no explanation why
this should be so has been suggested. This is not the only unsatisfactory
feature in this picture. Notice that since gauge singlets $S$'s are present,
we should include terms like $S\Psi\bar\Psi$ and $SSS$ in the superpotential.
The whole analysis in Ref.\ \rnath\ loses its validity as a result.
Furthermore, it remains a mystery why a third VEV along the direction of $e^+$
should not develop. Finally, any discrete symmetry will be broken by the four
VEV's in\ \evev. Thus this mechanism is not compatible with Ref.\ \rlthiggs.

In order to avoid these undesirable features, we will examine carefully the
roles the singlets and the discrete symmetry ought to play in the symmetry
breaking. We find that although mirror particles are still necessary, other
problems can be solved because the superpotential does not have to have flat
directions. Instead, it has a few isolated vacua corresponding to various low
energy gauge groups, including, of course, the standard model.

\newsec{The New Breaking Mechanism of \su33}

Consider the general \su33 model with an additional discrete symmetry
$C_N$ (The model in Ref.\ \roxford\ is thus a special case.). The
superpotential obeys
\eqn\esp{\eqalign{W\ =&\ f^{abc}_{ABC}\Psi_a^A\Psi_b^B\Psi_c^C+
  {f^\prime}^{abc}_{ABC}\bar\Psi_a^A\bar\Psi_b^B\bar\Psi_c^C\cr
&+g^{abc}_{ABC}\Psi_a^A\bar\Psi_b^BS_c^C+m^{ab}_{AB}\Psi_a^A\bar\Psi_b^B\cr
 &+h^{abc}_{ABC}S_a^AS_b^BS_c^C+M^{ab}_{AB}S_a^AS_b^B
  +{M^\prime_A}^2S_0^A,\cr
 &+({\rm non-renormalizable\ \ terms})\cr}}
where indices $a$ and $b$ indicate the discrete charges of the fields under
$C_N$ (For example, $\Psi_a$ is the field which transforms like $i_N^a$, where
$i_N$ is the N'th root of $1$.), $\Psi$ and $\bar\Psi$ stand for $\Psi_\ell$
and $\bar\Psi_\ell$ (We have omitted any term containing $\Psi_{L(R)}$ or
$\bar\Psi_{L(R)}$ because they are not relevant to the symmetry breaking
mechanism we are considering.), and indices A,B,C specify different
generations of fields with the same quantum numbers. The term $\Psi\Psi\Psi$
stands for $\epsilon_{i_1i_2i_3}\epsilon^{j_1j_2j_3}
\Psi^{i_1}_{j_1}\Psi^{i_2}_{j_2}\Psi^{i_3}_{j_3}$, where the $i$'s and $j$'s
are $SU(3)_L$ and $SU(3)_R$ indices respectively, and likewise for
$\bar\Psi\bar\Psi\bar\Psi$. Similarly the term $\Psi\bar\Psi$ is a shorthand
for $\Psi^i_j\bar\Psi^j_i$. Since in most cases $M^\prime_A$'s can be set to
zero by a shift of $S_0^A$'s, we will drop these terms from now on. Notice
that these coefficients are symmetric under permutations of indices. Also,
since $C_N$ is a symmetry of the theory, each term in $W$ should
carry no discrete charge. Therefore any coefficient vanishes if its $C_N$
indices do not sum to zero.

To obtain the ground state we minimize both the $F$-term and the $D$-term. In
order not to break SUSY at this stage, we would like to solve
\eqn\efterm{{\partial W \over \partial \phi}_{|_{\phi=S,\Psi,\bar\Psi}}=0}
\centerline{and}
\eqn\edt{\sum_{\phi=\Psi}\phi^\dagger T^a_L\phi-
  \sum_{\phi=\bar\Psi}\phi T^a_L \phi^\dagger=
 \sum_{\phi=\Psi}\phi T^a_R\phi^\dagger-
 \sum_{\phi=\bar\Psi}\phi^\dagger T^a_R \phi=0,}
where $T^a_{L(R)}$'s are the generators of $SU(3)_{L(R)}$. Obviously the
origin is always a solution for these equations. Since there are as many
equations in\ \efterm\  as there are variables, naively we would expect the
additional constraints of\ \edt\ to exclude other solutions. To show that this
is not the
case, let's write down those equations in\ \efterm\ more carefully,
\eqna\eft
$$\eqalignno{3{f^\prime}^{abc}_{ABC}\langle\bar\Psi_a^A\rangle
  \langle\bar\Psi_c^C\rangle
  +&\langle\Psi_a^A\rangle(g^{ab0}_{ABC}\langle S_0^C\rangle+m_{AB}^{ab})
  +g^{abc}_{ABC}\langle\Psi_{a}^A\rangle\langle S_{c\ne0}^C\rangle\cr
 &\ =\ {\partial W \over \partial\bar\Psi_b^B}\ =\ 0,&\eft a\cr
 &\cr
 3f^{abc}_{ABC}\langle\Psi_b^B\rangle \langle\Psi_c^C\rangle
  +&\langle\bar\Psi_b^B\rangle(g^{ab0}_{ABC}\langle S_0^C\rangle+m_{AB}^{ab})
  +g^{abc}_{ABC}\langle\bar\Psi_{b}^B\rangle\langle S_{c\ne0}^C\rangle\cr
 &\ =\ {\partial W \over \partial\Psi_{a}^A}\ =\ 0,&\eft b\cr
 &\cr
 g^{ab0}_{ABC}\langle\Psi_a^A\rangle \langle\bar\Psi_{b}^B\rangle
  +&3h^{ab0}_{ABC}\langle S_a^A\rangle\langle S_{b}^B\rangle
  +2M_{AC}^{00}\langle S_0^A\rangle\cr
 &\ =\ {\partial W \over \partial S_{0}^C}\ =\ 0,&\eft c\cr
 &\cr
 g^{abc}_{ABC}\langle\Psi_a^A\rangle \langle\bar\Psi_{b}^B\rangle
  +&3h^{abc}_{ABC}\langle S_a^A\rangle\langle S_{b}^B\rangle
  +2M_{AC}^{ac}\langle S_{a}^A\rangle\cr
 &\ =\ {\partial W \over \partial S_{c\ne0}^C}\ =\ 0.&\eft d\cr}$$
If we set each term in\ \eft{a,b}\ to be zero individually, even
though the number of constraints seems to increase, many of them may be in
fact degenerate, thus there may be less independent constraints than
equations in\ \eft{a,b}. So we replace\ \eft{a,b}\ with the following,
\eqna\efta
$$\eqalignno{{f^\prime}^{ab}_{ABC}\langle\bar\Psi_a^A\rangle
  \langle\bar\Psi_b^B\rangle\ &=\ 0,&\efta a\cr&\cr
 f^{abc}_{ABC}\langle\Psi_a^A\rangle \langle\Psi_b^B\rangle\ &=\ 0,
 &\efta b\cr&\cr
 g^{abc}_{ABC}\langle\Psi_a^A\rangle\langle S_{c\ne0}^C\rangle\ &=\ 0,
 &\efta c\cr&\cr
 g^{abc}_{ABC}\langle\bar\Psi_{b}^B\rangle\langle S_{c\ne0}^C\rangle
 \ &=\ 0,&\efta d\cr&\cr
 g^a_{ABC}\langle S_0^C\rangle+m_{AB}^a\ &=\ 0.&\efta e\cr}$$
Eq.\ \efta{e}\ holds if at least one $\langle\Psi_a^A\rangle$ or
$\langle\bar\Psi_{-a}^B\rangle$ is nonzero, which is exactly the type of
solution we are looking for. Note that eq.\ \efta{e}\ is a single non-matrix
constraint (in gauge space), even though it is derived from\ \eft{a,b}, two
matrix constraints. This can represent a vast decrease in the number of
constraints. Also, note that nonzero $\langle\Psi_a^A\rangle$ and
$\langle\bar\Psi_{-a}^B\rangle$, together with eq.\ \efta{c,d}, generally
imply $\langle S_{b\ne0}^C\rangle\ =\ 0$.

Although we have already greatly simplified these equations, the solutions can
still be very complicated, especially if the number of generations is large.
For example, we have found that solutions which break the hypercharge $U(1)$
can exist if $n_{S_0}\ge 5$, where $n_{S_0}$ is the number of $S_0$'s. We will
not consider such cases. More interesting is when $n_{S_0}$ is small. For
$n_{S_0}\le 4$, it is often possible to break $SU(3)^3$ to $SU(3)\times
SU(2)\times U(1)$ and also leave an unbroken combination of $C_N$ with an
element of the gauge group. This unbroken symmetry can preserve the lightness
of the Higgs doublets\rlthiggs. As an example, we will consider the simplest
case, which corresponds to $n_{S_0}=2$ and $n_{\Psi_{a_1}}=
n_{\bar\Psi_{-a_1}}= n_{\Psi_{a_2}}=n_{\bar\Psi_{-a_2}}=1$, where
$n_{\Psi_{a}}$ and $n_{\bar\Psi_{-a}}$ are the number of $\Psi_a$'s and
$\bar\Psi_{-a}$'s respectively. Applying the constraints of\ \edt,\ \eft{c,d}\
and\ \efta{a,b,e}, we find that, after a choice of basis, the solutions are of
the following form,
$$\langle\Psi_{a_1}\rangle\ =\ \left(\matrix{0&0&0\cr 0&0&0\cr
0&0&v\cr}\right),\indent \langle\bar\Psi_{-a_1}\rangle\ =\
\left(\matrix{0&0&0\cr
0&0&0\cr 0&0&v^*\cr}\right),$$
$$\langle\Psi_{a_2}\rangle\ =\ \left(\matrix{0&0&0\cr 0&0&0\cr
0&w&0\cr}\right),\indent \langle\bar\Psi_{-a_2}\rangle\ =\
\left(\matrix{0&0&0\cr
0&0&w^*\cr 0&0&0\cr}\right),$$
Notice that the VEV's of $\Psi_a$ and $\bar\Psi_{-a}$ are hermitian conjugate
to each other and that $\langle\Psi_{a_1}\rangle$ is perpendicular to
$\langle\Psi_{a_2}\rangle$. The magnitude of $v$ and $w$ is determined and is
related to the characteristic mass scale in the original Lagrangian,  {\it
i.e.} $v,w\sim s$, the supposed string scale, if they are not zero. This is
compatible with the result from renormalization group calculation of the
running coupling constants in certain versions of the non-minimal SUSY SM\
\rlthiggs.

Notice that the transformation properties of these VEV's under $C_N$ are
exactly such that the product of $C_N$ and a certain element of \su33 remains
unbroken. In other words, the VEV's in the $\bar\Psi$'s do not further break
the symmetry. Thus we have justified one of the assumptions made in Ref.\
\rlthiggs.

Since there are three degenerate vacua : the origin, the one with $v\ne0,w=0$
, and the one with both $v,w\ne0$, we have to determine the true vacuum by
soft SUSY breaking terms. Whether or not it favors the vacuum we want over the
others depends on the form of the soft breaking term and the coefficients in
the superpotential, which are unknown. Nevertheless it seems likely that the
standard model is favored.

Compared to the model in Ref.\ \rlthiggs, we have introduced new particles
$S$'s. After the symmetry breaking, all of them gain masses of the same order
as $v$ and $w$. Therefore they are invisible at low energy.

\newsec{Conclusion}

We have detailed a supersymmetric gauge model with a gauge symmetry breaking
mechanism very similar to SUSY $SU(5)$. In fact, almost everything good about
$SU(5)$ can be carried over to our model, while it offers a few additional
nice features of its own. To name a few, not only it is a likely product of
string theory, but also it offers a natural way to solve the hierarchy problem
and, perhaps, the strong $CP$ problem\ \ref\rself{E.D. Carlson and M.Y. Wang,
Harvard Preprint HUTP-92/A057; for a supersymmetric version, see ``M.Y. Wang
and E.D. Carlson, Harvard Preprint HUTP-93/A004".}. By natural we mean that
neither fine-tuning of continuous parameters nor introduction of exotic
particles is necessary. However, a few assumptions have been made in the
course of our argument, reflecting our ignorance concerning issues such as
what is happening at the Planck scale and what breaks SUSY\
\ref\rselfagain{M.Y. Wang and W.A. Leaf, Preprint in preparation.}. It is
unlikely that these remaining questions will be completely understood in the
near future. Nevertheless, we are encouraged to see that in principle the
long-standing hierarchy problem can be solved with the right choice of
discrete parameters within the traditional theoretical framework without
invoking new revolutionary concepts.

\bigbreak\bigskip\bigskip\centerline{{\bf Acknowledgements}}\nobreak
We thank S.Coleman and M.Golden for their suggestions. This research is
supported in part by the National Science Foundation under grant
\#PHY--87--14654, and the Texas National Research Laboratory Commission under
Grant \#RGFY9206.

\vfill\eject
\listrefs
\bye